\documentstyle[12pt]{article}

\input{feynman}

\advance\textheight by \topskip

\newcommand{\mpi}{m_{\pi}}
\newcommand{\cl}[1]{{\cal #1}}
\newcommand{\lat}{\la^{(2)}_{\rm even}}
\newcommand{\bma}[1]{\left( \begin{array}{#1}}
\newcommand{\ema}{\end{array} \right)}
\newcommand{\hmu}{^{\mu}}

\newcommand{\lmu}{_{\mu}}
\newcommand{\lnu}{_{\nu}}
\newcommand{\be}{\begin{equation}}
\newcommand{\ee}{\end{equation}}
\newcommand{\bea}{\begin{eqnarray}}
\newcommand{\eea}{\end{eqnarray}}
\newcommand{\eq}[1]{eq.~(\protect\ref{#1})}
\newcommand{\nn}{\nonumber}
\newcommand{\kruis}{^{\dagger}}
\newcommand{\lh}{\Bigl[}
\newcommand{\rh}{\Bigr]}
\newcommand{\lrh}{\Bigl(}
\newcommand{\rrh}{\Bigr)}
\newcommand{\grootlh}{\Biggl[}
\newcommand{\grootrh}{\Biggr]}

\newcommand{\lov}{\la_{\rm odd}^{(4)}}

\newcommand{\gev}{\; {\rm GeV}}
\newcommand{\mev}{\; {\rm MeV}}
\newcommand{\pijl}{\rightarrow}
\newcommand{\lb}[1]{\label{#1}}
\newcommand{\la}{{\cal L}}
\newcommand{\ii}{{\rm i}}
\newcommand{\nc}{N_{\rm c}}
\newcommand{\trace}{{\rm Tr}}
\newcommand{\eps}{\epsilon}
\newcommand{\half}{\frac{1}{2}}
\newcommand{\fp}{f_{\pi}}
\begin{document}

\title{Cross section of $\gamma\gamma\pijl\pi^+\pi^-\pi^0$
in chiral perturbation theory}

\author{J.W. Bos\thanks
{e-mail jwb@joule.phy.ncu.edu.tw}, Y.C. Lin, and H.H. Shih\\
Department of Physics,
National Central University \\
Chung Li, Taiwan 32054 \\
\\
NCUHEP 94-04}

\date{}

\maketitle

\begin{abstract}
We give the amplitude for $\gamma\gamma\pijl\pi^+\pi^-\pi^0$
in leading order chiral perturbation theory.
For the case of real photons we calculate the total and
differential cross section.
Furthermore, we give the dependence of the total
cross section on the invariant mass of one of the
photons.
\end{abstract}

Chiral perturbation theory (ChPT) \cite{li71,wein79,gass84} has proven
to be an useful method to describe reactions involving low-energy
mesons.
The method is closely related to QCD, and therefore comparison
of ChPT predictions with experimental data provides an indirect test
for QCD.
Reactions that are particularly useful for this purpose are
inelastic two photon processes.
These type of reactions
attracted a lot of experimental attention recently.
In this letter we calculate the cross section for the specific
channel $\gamma\gamma\pijl\pi^+ \pi^-\pi^0$ in leading
order ChPT.
In contrast to the $\gamma\gamma\pijl\pi^+\pi^-$ channel
this channel is sensitive to the {\em anomalous\/} or
{\em odd intrinsic parity\/} sector of QCD.

In the ChPT approach (we need to consider here the meson sector only),
one starts from the most general gauge invariant effective lagrangian
in terms of meson degrees of freedom, which reflects the approximate
chiral ${\rm SU}(3)_L \times {\rm SU}(3)_R$ symmetry of the QCD
lagrangian.
Based on this effective lagrangian a perturbation scheme is developed
in which one expands in a systematic way in the external momenta and
the meson masses.
The leading terms in the expansion of an amplitude in these variables
are given by the tree-level diagrams with vertices from the leading order
lagrangian.
Loop diagrams, and tree-level diagrams containing vertices from
the next to leading order lagrangian, are suppressed by a power
$(p/\Lambda_{\chi})^2$ where $p$ denotes the four momenta or mass of
the mesons, and $\Lambda_{\chi}$ is the scale in the energy expansion,
$\Lambda_{\chi} \approx 1 \gev$.
At low energies, and since the meson masses are small compared to
$\Lambda_{\chi}$, one needs to keep only the first terms in this
expansion. We will here only consider the amplitude in leading order,
i.e.\ the tree-level.

The general ChPT lagrangian in the meson sector can be divided
into an {\em even\/} and
an {\em odd\/} intrinsic parity part \cite{witt83}.
The first consists of interactions among only an even number of mesons
while the second consists of interactions among only an odd number of
mesons.
The  effective lagrangian of ChPT in the odd intrinsic parity
sector reads
\be\lb{al}
	\la_{\rm odd}=\la^{(4)}_{\rm odd}+\la^{(6)}_{\rm odd}+\ldots,
\ee
where the superscripts denote the momentum dimension.
The leading term  in \eq{al}, $\lov$, is the well-known anomalous
Wess-Zumino-Witten term~\cite{witt83,wess71}.
For our application it is sufficient to take
\bea\lb{la4}
	\la^{(4)}_{\rm odd}&=&-\frac{\ii\nc e}{48\pi^2}
	\eps^{\mu\nu\alpha\beta}A\lmu\trace\lh Q(\partial\lnu\Sigma
	\partial_{\alpha}\Sigma\kruis\partial_{\beta}\Sigma
	\Sigma\kruis-\partial\lnu\Sigma\kruis
	\partial_{\alpha}\Sigma\partial_{\beta}\Sigma\kruis
	\Sigma))\rh
\nn\\
	&&\mbox{}-\frac{\ii\nc e^2}{24\pi^2}\eps^{\mu\nu\alpha\beta}
	(\partial\lmu A\lnu )A_{\alpha}\trace\lh Q^2\partial_{\beta}
	\Sigma\Sigma\kruis +Q^2\Sigma\kruis\partial_{\beta}\Sigma
\nn\\
	&&\mbox{}+\half Q\Sigma Q\Sigma\kruis\partial_{\beta}\Sigma\Sigma\kruis
	-\half Q\Sigma\kruis Q\Sigma\partial_{\beta}\Sigma\kruis\Sigma\rh.
\eea
The meson fields are contained in the SU(3) matrix $\Sigma$
given by
\be
	\Sigma =\exp (2\ii\pi/\fp),
\ee
where
\be
	\pi =\frac{1}{\sqrt{2}}\bma{ccc}\frac{1}{\sqrt{2}}\pi^0+
	\frac{1}{\sqrt{6}}\eta &\pi^+&K^+\\\pi^-&-\frac{1}{\sqrt{2}}\pi^0+
	\frac{1}{\sqrt{6}}\eta &K^0\\K^-&\bar{K^0}&-\frac{2}{\sqrt{6}}\eta\ema
\ee
and $\fp$ is the pion decay constant, $\fp=94\mev$.
In \eq{la4} $A\lmu$ is the electromagnetic field and $Q$ is the charge
operator in flavor space given by
\be
	Q={\rm diag}(2/3,-1/3,-1/3).
\ee
The second term in the lagrangian \eq{al}, $\la^{(6)}_{\rm odd}$,
is of higher order in the momentum and mass expansion
\cite{bijn91} and therefore omitted in this leading order calculation.

The effective lagrangian in the even intrinsic parity sector
of ChPT is, to leading order, given by
\be\lb{la2}
	\la^{(2)}_{\rm even}=\frac{\fp}{4}\trace
	(D\lmu\Sigma D\hmu\Sigma\kruis+
	\chi\Sigma\kruis + \Sigma\chi\kruis),
\ee
where the covariant derivative, $D\lmu$, is defined by
\be
	D\lmu\Sigma =\partial\lmu\Sigma +\ii eA\lmu (Q\Sigma -\Sigma Q)
\ee
and $\chi$ is the chiral symmetry breaking mass matrix,
\be
\chi=B\,{\rm diag}(m_u,m_d,m_s).
\ee

We now turn to the reaction $\gamma\gamma\pijl\pi^+\pi^-\pi^0$ using
the lagrangian \eq{la4} and \eq{la2}.
In fig.~1 we define our notation for the external momenta,
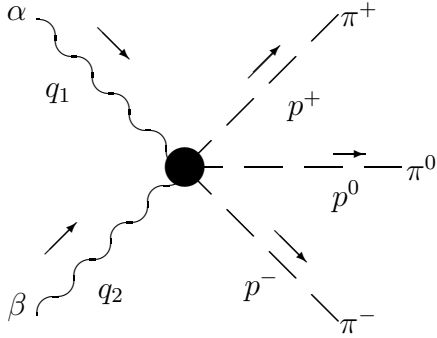
\begin{figure}[t]
\centerline{
\begin{picture}(15000,15000)
\drawline\photon[\NE\CURLY](0,2000)[8]
\put(\pfrontx,\pfronty){\makebox(0,0)[br]{$\beta\ $}}
\addtolength{\pmidy}{-2000}\put(\pmidx,\pmidy){\makebox(0,0){$q_2$}}
\addtolength{\pmidy}{1500}\addtolength{\pmidx}{-2500}
\put(\pmidx,\pmidy){\vector(1,1){1200}}
\drawline\scalar[\SE\REG](\pbackx,\pbacky)[4]
\put(\pbackx,\pbacky){\makebox(1500,0){$\pi^-$}}
\addtolength{\pmidy}{-1000}
\put(\pmidx,\pmidy){\makebox(0,0)[t]{$p^-$}}
\addtolength{\pmidx}{500}\addtolength{\pmidy}{1500}
\put(\pmidx,\pmidy){\vector(1,-1){1200}}
\drawline\scalar[\E\REG](\pfrontx,\pfronty)[4]
\put(\pbackx,\pbacky){\makebox(1500,0){$\pi^0$}}
\addtolength{\pmidx}{2000}\addtolength{\pmidy}{-500}
\put(\pmidx,\pmidy){\makebox(0,0)[t]{$p^0$}}
\addtolength{\pmidy}{1000}\addtolength{\pmidx}{-500}
\put(\pmidx,\pmidy){\vector(1,0){1200}}
\drawline\photon[\NW\CURLY](\pfrontx,\pfronty)[8]
\put(\pbackx,\pbacky){\makebox(0,0)[br]{$\alpha\ $}}
\addtolength{\pmidx}{-2000}\put(\pmidx,\pmidy){\makebox(0,0){$q_1$}}
\addtolength{\pmidx}{1500}\addtolength{\pmidy}{2500}
\put(\pmidx,\pmidy){\vector(1,-1){1200}}
\drawline\scalar[\NE\REG](\pfrontx,\pfronty)[4]
\addtolength{\pmidx}{1000}\put(\pmidx,\pmidy){\makebox(0,0)[tl]{$p^+$}}
\addtolength{\pmidx}{-1500}\addtolength{\pmidy}{500}
\put(\pmidx,\pmidy){\vector(1,1){1200}}
\put(\pfrontx,\pfronty){\circle*{1500}}
\put(\pbackx,\pbacky){\makebox(1500,0)[r]{$\pi^+$}}
\end{picture}}
\caption{\em Kinematics of the $\gamma\gamma\pijl\pi^+\pi^-\pi^0$ process}
\end{figure}
and the Feynman diagrams contributing to the leading order
amplitude are shown in fig.~2.
\begin{figure}[t]
\centerline{
\begin{picture}(30000,24000)
\put(0,0){\begin{picture}(15000,14000)
\drawline\photon[\NE\CURLY](0,2000)[6]
\drawline\scalar[\SE\REG](\pbackx,\pbacky)[3]
\drawline\scalar[\E\REG](\pfrontx,\pfronty)[3]
\drawline\photon[\NW\CURLY](\pfrontx,\pfronty)[6]
\drawline\scalar[\NE\REG](\pfrontx,\pfronty)[3]
\put(\pfrontx,0){\makebox(0,0){$(\rm c)$}}\end{picture}}
\put(0,14000){\begin{picture}(15000,12000)
\drawline\photon[\NE\CURLY](0,2000)[6]
\drawline\scalar[\SE\REG](\pbackx,\pbacky)[3]
\drawline\scalar[\NE\REG](\pfrontx,\pfronty)[3]
\drawline\scalar[\N\REG](\pfrontx,\pfronty)[3]
\addtolength{\pmidx}{-1000}
\put(\pmidx,\pmidy){\makebox(0,0)[r]{$\pi^+,\,\pi^-$}}
\drawline\photon[\NW\CURLY](\pbackx,\pbacky)[6]
\drawline\scalar[\NE\REG](\pfrontx,\pfronty)[3]
\put(\pfrontx,0){\makebox(0,0){$(\rm a)$}}\end{picture}}
\put(15000,14000){\begin{picture}(15000,12000)
\drawline\photon[\NE\CURLY](0,2000)[6]
\drawline\photon[\NW\CURLY](\pbackx,\pbacky)[6]
\drawline\scalar[\E\REG](\pfrontx,\pfronty)[3]
\addtolength{\pmidy}{1000}\put(\pmidx,\pmidy){\makebox(0,0){$\pi^0,\,\eta$}}
\put(\pmidx,0){\makebox(0,0){$(\rm b)$}}
\drawline\scalar[\SE\REG](\pbackx,\pbacky)[3]
\drawline\scalar[\E\REG](\pfrontx,\pfronty)[3]
\drawline\scalar[\NE\REG](\pfrontx,\pfronty)[3]\end{picture}}
\end{picture} }
\caption{\em Leading order Feynman diagrams for
$\gamma\gamma\pijl\pi^+\pi^-\pi^0$.}
\end{figure}
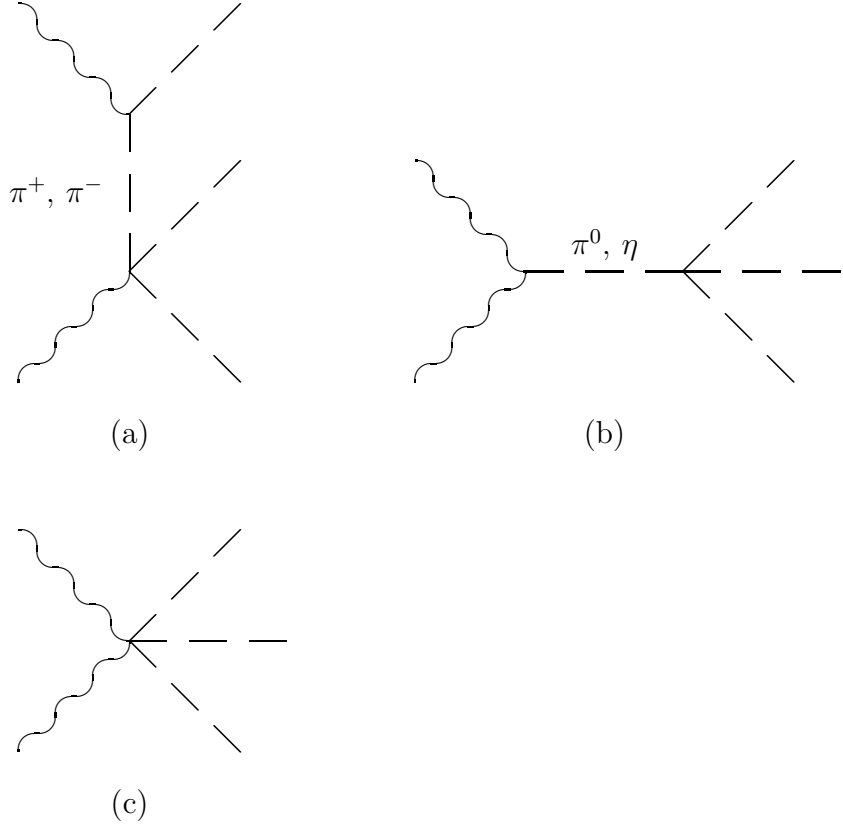
The pole diagrams (a) and (b) in fig.~2 consist of one vertex from the odd
intrinsic parity lagrangian $\lov$ and one from the normal intrinsic
parity lagrangian $\lat$, while diagram (c) consist only of
one vertex from  $\lov$.
Since the $\eta\pi^+\pi^-\pi^0$ coupling is an isospin breaking
effect, diagram (b) with an intermediate $\eta$ propagator is suppressed.

Applying the standard Feynman rules and averaging over the
polarizations of the initial photons the amplitude squared reads
\be
	|\cl{M}|^2 = \frac{1}{4}\Gamma_{\alpha\beta}\Gamma^{\alpha\beta},
\ee
where
\bea\lb{amplitude}
	\Gamma^{\alpha\beta}&=&\frac{\ii e^2\nc}
	{12\pi^2 \fp^3}\grootlh\eps^{\alpha\beta\mu\nu}{q_1}\lmu{q_2}\lnu\lrh
	\frac{(p^++p^-)^2-\mpi^2}{(q_1+q_2)^2-\mpi^2}-\frac{1}{3}\rrh
\nn\\*
	&&\mbox{}+\eps^{\alpha\beta\mu\nu}\lrh\frac{4}{3}{q_1}\lmu{q_2}\lnu
	-(q_1-q_2)\lmu (p^++p^-)\lnu\rrh
\nn\\*
	&&\mbox{}-\lh\eps^{\alpha\mu\nu\rho}\lrh\frac{(2p^+-q_2)^{\beta}}
	{q_2^2-2p^+\cdot q_2}p^-\lmu {q_1}_{\nu}(p^+-q_2)_{\rho}
\nn\\*
	&&\mbox{}+\frac{(2p^--q_2)^{\beta}}
	{q_2^2-2p^-\cdot q_2}p^+\lmu {q_1}_{\nu}(p^--q_2)_{\rho}\rrh
	+(q_1\leftrightarrow q_2,\, \alpha\leftrightarrow\beta)\rh\grootrh.
\eea
This is in agreement with the early result by
Adler et al \cite{adle71} using a current algebra
approach\footnote{To arrive at eq.~(\protect\ref{amplitude})
one must take for $F^{\pi}$,
$F^{3\pi}$, and $x$ in Ref.~\cite{adle71}
\be
F^{\pi}=\frac{\alpha}{\pi\fp};\,\,eF^{3\pi}=\frac{\alpha}{\pi\fp^3} ;\,\,x=0.
\ee}.

The differential cross section for $\gamma\gamma\pijl\pi^+\pi^-\pi^0$
is for real photons given by
\be
	d\sigma = \frac{1}{2E_{\rm cms}}|\cl{M}|^2\frac{d\vec{p}^{\,+}}
	{2E^+(2\pi)^3}\frac{d{\vec{p}}^{\,\,0}}{2E^0(2\pi)^3}
	\frac{d\vec{p}^{\,\,-}}{2E^-(2\pi)^3}(2\pi)^4
	\delta^4(q_1+q_2-p^+-p^0-p^+),
\ee
where the momenta are the same as in fig.~1, and $E_{\rm cms}$ is
the total center of mass energy of the two photons.
The amplitude squared depends on five kinematical variables
for which we choose the invariants
\bea
	&&s=(q_1+q_2)^2;\, s_1=(p^++p^0)^2;\, s_2=(p^-+p^0)^2
\nn\\*
	&&t_1=(q_1-p^+)^2;\, t_2=(q_2-p^-)^2.
\eea
The total cross section, $\sigma$, as a function of $s=E_{\rm cms}^2$,
is given in terms of these invariants by
\be\lb{sig}
	\sigma(s)=\frac{\pi}{32s^2(2\pi)^5}\int\!\!\frac{|\cl{M}|^2}
	{\sqrt{-\Delta_4}}\, dt_1dt_2ds_1ds_2,
\ee
where $\Delta_4=\Delta_4(s,s_1,s_2,t_1,t_2)$ is given by the
determinant
\be
	\Delta_4=\frac{1}{16}\det\bma{cccc}2s_2&s_2-t_1&s+s_2-\mpi^2&s_2\\
	s_2-t_1&0&s&\mpi^2-t_2\\
	s+s_2-\mpi^2&s&2s&s-s_1+\mpi^2\\
	s_2&\mpi^2-t_2&s-s_1+\mpi^2&2\mpi^2\ema.
\ee
The range of the integration in \eq{sig} is the physical region of
the process, which satisfies $\Delta_4\le 0$ \cite{byck73}.

In fig.~3-a we show our prediction for the total
\begin{figure}[t]
\vspace{1cm}
\caption{\em (a) Total and (b) differential cross section for
different values of $s$ (units of GeV).}
\end{figure}
cross section as a function of the total center of mass energy
$E_{\rm cms}=\sqrt{s}$, where we varied $E_{\rm cms}$ from threshold,
$E_{\rm cms}=3\mpi$
up to $E_{\rm cms}=0.7\gev$.
Of course, at and above this point contributions from resonances
like the $\eta$ and $\eta'$, and also loop diagrams are expected
to be important.
More essential features of the reaction mechanism are contained
in the differential cross sections.
Fig.~2-b shows $d\sigma/ds_1$ as a
function of the invariant mass of the final
$\pi^+\pi^0$ system, $\sqrt{s_1}$,
for different values of $s$. The range of $\sqrt{s_1}$ is determined
by the available phase space.
Because of charge conjugation invariance the differential
cross section $d\sigma/ds_2$ has the same form as $d\sigma/ds_1$ and
is therefore not shown here.

In a electron-positron collider, a measurement with two real photons
requires double tagging of the final electron and positron
at $0^{\circ}$ (for discussion of experimental aspects of photon photon
reactions see e.g.\ Ref.~\cite{berg87}). The detection efficiency
is increased by allowing for virtual photons, $q_i^2 \ne 0$,
as in a single tagging or no tagging experimental situation.
With virtual photons \eq{sig} for the total cross section
becomes
\be
	\sigma(s)=\frac{\pi}{32\lambda(s,q_1^2,q_2^2)
	(2\pi)^5}\int\!\!\frac{|\cl{M}|^2}
	{\sqrt{-\Delta_4}}\, dt_1dt_2ds_1ds_2,
\ee
where $\Delta_4=\Delta_4(s,s_1,s_2,t_1,t_2,q_1^2,q_2^2)$ is
given by the determinant
\be
	\Delta_4=\frac{1}{16}\det\bma{cccc}2s_2&s_2-t_1+q_2^2&s+s_2-\mpi^2&s_2\\
	s_2-t_1+q_2^2&2q_2^2&s-q_1^2+q_2^2&q_2^2+\mpi^2-t_2\\
	s+s_2-\mpi^2&s-q_1^2+q_2^2&2s&s-s_1+\mpi^2\\
	s_2&q_2^2+\mpi^2-t_2&s-s_1+\mpi^2&2\mpi^2\ema
\ee
and
\be
	\lambda(s,q_1^2,q_2^2) = (s-q_1^2-q_2^2)-4q_1^2q_2^2.
\ee
Here we will consider the situation of a single tagging
experiment, i.e.\ only one of the initial photons is real, $q_1^2=0$, while
the other is virtual, $q_2^2\ne 0$.
The dependence of the total
cross section on the virtual photon mass $W_q=\sqrt{q_2^2}$, with
$q_1^2=0$, is shown in fig.~4.
\begin{figure}[t]
\vspace{1cm}
\caption{\em $q^2$ dependence of the total cross section for
different values for $s$ (units of GeV)}
\end{figure}
Again, the different curves correspond to different values of $s$.
We varied $W_q$ from $W_q = 0$ up to the $W_q = 2\mpi$,
the threshold for the production of two on-shell pions by the
virtual photon.
The dependence of the total cross section
on $W_q$ is moderate, only a variation
of about 10 \% over the whole range of  $W_q$.
It indicates that compared to a double tagging situation
the cross section in the experimentally more favorable
single tagging situation is not much smaller in the
kinematical range considered here.

The energy range for which the above calculation can be
used depends on the magnitude of next-to-leading order corrections.
This calculation is only valid in leading order in the
$(p/\Lambda_{\chi})^2$ expansion.
In next to leading order ChPT not only chiral loops
but also the {\em free\/} parameters from $\la_{\rm odd}^{(6)}$
in \eq{al} enter the description.
To give an estimate of higher order effect it is therefore
necessary to determine these parameters, which usually cannot
be done in a model independent way.
However, comparing with
similar two photon processes
it has been shown that the width of $\pi^0$ decay
into two photons is described very well
by a leading order calculation \cite{dono85,dono89}. Since this process
is sensitive to the same part of the ChPT lagrangian as
$\gamma\gamma\pijl\pi^+\pi^-\pi^0$ we expect that also in the latter
higher order correction are small.
For virtual photons corrections to $\pi^0$ decay
from ${\cal O}(p^6)$ terms have shown to be sizable \cite{bijn93},
and we therefore expect that this goes through for
photon photon collisions involving virtual photons.
A detailed study of the importance of higher order effects
is currently in progress.

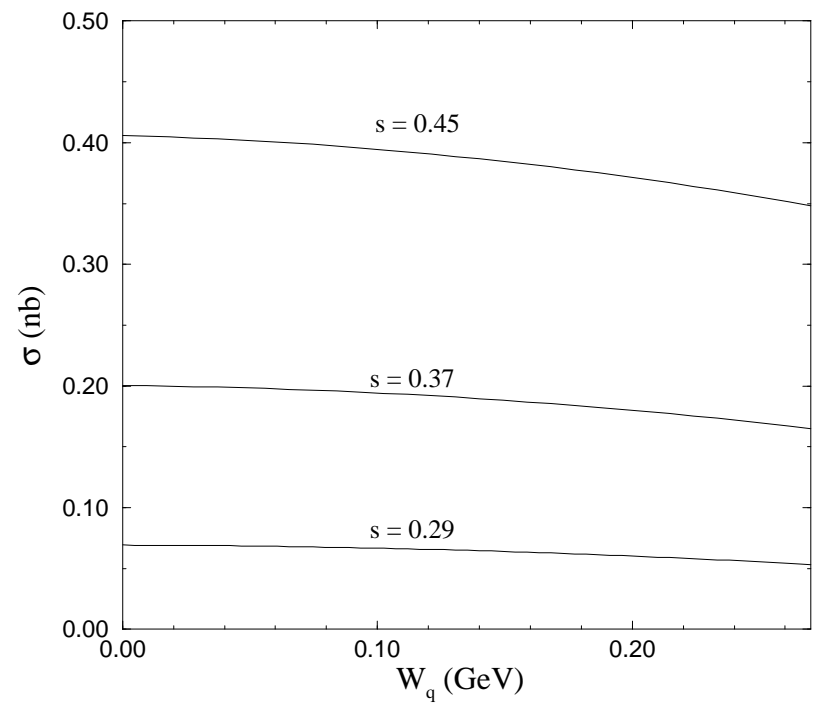
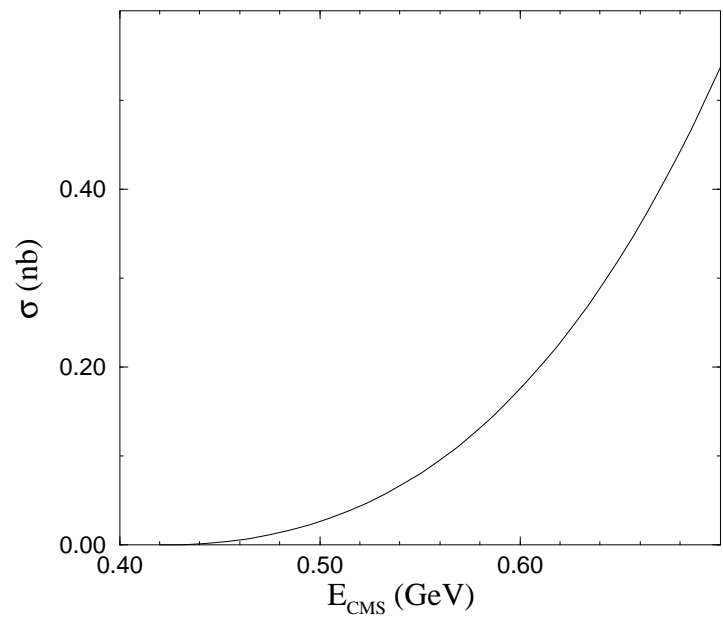
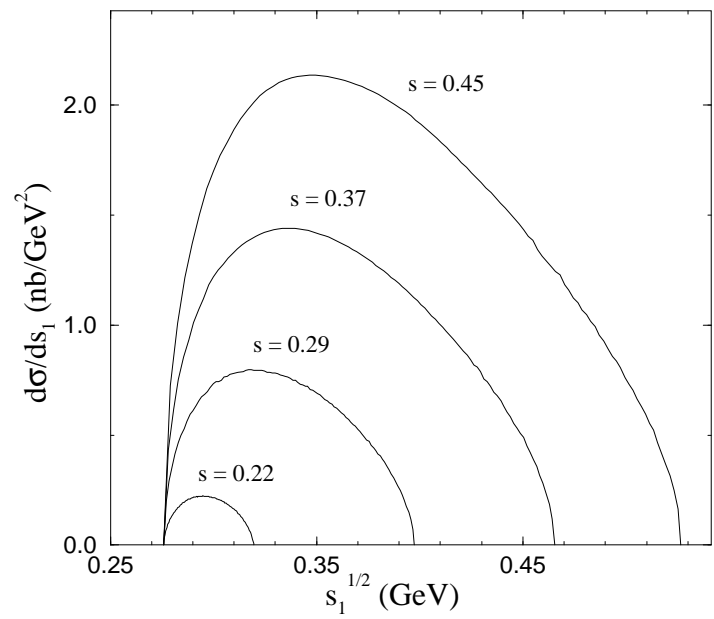
\end{document}